\documentclass[preprint,showpacs,preprintnumbers,amsmath,amssymb]{revtex4}

\usepackage{graphicx}
\usepackage{dcolumn}
\usepackage{bm}

\begin{document}


\title{Baryonium, Tetra-quark State and Glue-ball in Large $N_c$ QCD}

\author{Chun Liu}
\affiliation{Institute of Theoretical Physics,
Chinese Academy of Sciences,\\
P.O. Box 2735, Beijing 100080, China}
 \email{liuc@itp.ac.cn}

\date{\today}

\begin{abstract}
From the large-$N_c$ QCD point of view, baryonia, tetra-quark states, 
hybrids, and glueballs are studied.  The existence of these states is 
argued for.  They are constructed from baryons.  In $N_f=1$ large $N_c$ 
QCD, a baryonium is always identical to a glueball with $N_c$ valence 
gluons.  The ground state $0^{-+}$ glueball has a mass about $2450$ MeV.  
$f_0(1710)$ is identified as the lowest $0^{++}$ glueball.  The lowest 
four-quark nonet should be $f_0(1370)$, $a_0(1450)$, $K^*_0(1430)$ 
and $f_0(1500)$.  Combining with the heavy quark effective theory, 
spectra of heavy baryonia and heavy tetra-quark states are predicted.  
$1/N_c$ corrections are discussed.  
\end{abstract}



\maketitle

\section{Introduction}

The large-$N_c$ limit is one of the most important methods of
non-perturbative QCD \cite{0,0a,1}.  Properties of mesons can be
observed from the analysis of planar diagrams and baryons from
the Hartree-Fock picture.  While mesons are free, the interaction
between baryons is strong which scales as ${\mathcal O}(N_c)$.
This might imply a duality between the meson and baryon sectors 
\cite{1}.  This implication comes from the similarity between 
large-$N_c$ baryons and Polyakov-'t Hooft monopoles by taking 
$1/N_c$ as a coupling constant.  This duality was indeed partially 
realized \cite{2} in the Skyreme model \cite{3} in which a baryon is 
regarded as a soliton of a meson theory.

In this work, hadron spectra are described starting from baryons. 
Besides mesons and baryons, certain multiquark systems are also 
included.  Generally, color-singlet multiquark-gluon systems have 
been expected naively due to their colorlessness.  Now we try to 
give deeper reasoning about their existence and their description 
from the large-$N_c$ QCD point of view.  
In principle, for large-$N_c$ QCD, baryon theories exist
as dual ones to meson theories.  Such theories, however, are
strongly interacting, which are lack of a perturbative
description. Taking baryons as the starting point can be traced
back to Fermi and Yang \cite{4} long before the establishment of
QCD, while strange baryons were included by Sakita \cite{5}. Its
recent version can be found in Ref. \cite{6}.  Alternative to
taking baryons as basic building blocks, we will still use the
Hartree-Fock picture of large-$N_c$ baryons as Witten did \cite{1}
to do semi-quantitative analyses.

Baryons themselves have been studied in Refs. \cite{1,10,10a}.  
There are $N_c$ valence quarks in a baryon.  Baryon-baryon 
interactions are strong.  Molecular states of baryons can exist due 
to their strong interactions.  And they are just the nuclei \cite{1}.

\section{Baryonia}

Let us consider baryon-antibaryon systems.  The 
baryon-antibaryon system was mentioned in Ref. \cite{1}.  We 
will study its properties by assuming that it forms a bound state.  
The interaction between a baryon and an antibaryon can be as 
strong as that of baryon-baryon systems.  Therefore, we expect 
that molecular states of a baryon and an antibaryon also exist.  
Because of baryon-antibaryon annihilation, the baryon and 
anti-baryon in a baryonium is attractive at small distances, 
baryonia are more deeply bounded than nuclei.  

The relevant interactions can be classified into two cases.
The first one is that of glueball exchanges.  The antibaryon
inside a baryonium is not necessarily the anti-particle of the
baryon in the baryonium.  A possible example is the baryonium
composed of $\Delta^{++}(uuu)$ and 
$\bar{\Delta}^-(\bar{d}\bar{d}\bar{d})$ where the valence 
quark contents are given.  In this case, the interaction inside a 
baryonium is described by Fig. 1 when the two baryons are 
close enough.  Its $N_c$-dependence can be seen from the 
following.  Each gluon-quark vertex contributes $1/\sqrt{N_c}$.  
There are $N_c^2$ possible ways to make the first gluon.  
To make the second gluon, however, there are only $N_c$ 
possible ways, because the two gluons must compose a color 
singlet state.  Therefore, the interaction energy is still 
proportional to $N_c$.  Generally a baryonium mass is about 
$2N_c\Lambda_{\rm QCD}$.  The binding energy, though 
proportional to $N_c$, is expected to be smaller than baryon 
masses $\sim N_c\Lambda_{\rm QCD}$ because baryons are 
color singlet.  This is consistent with the molecular picture of 
baryonia.  In terms of the hadron language, the interaction 
is mediated by glueballs with the glueball-baryon coupling 
$\sim \sqrt{N_c}$.  Because of heaviness of glueballs, 
such a t-channel glueball exchange interaction might be 
suppressed unless it happens at small distances.  The 
short range interaction is also required by the confinement.  

\begin{figure}
\includegraphics{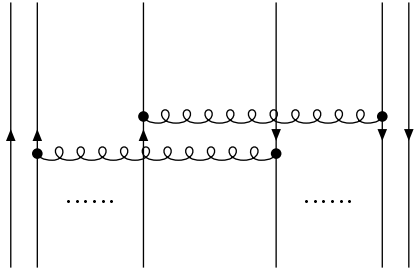}
\caption{\label{fig1}
Interaction inside a baryonium}
\end{figure}

The second case is that of meson exchanges.  When a quark and 
an antiquark have a common flavor in the baryonium, the 
interaction given by Fig. 2 \cite{1} plays a role.  This interaction is
regarded as a meson exchange.  Once any of the quarks is able to
annihilate with any of the antiquarks, namely, all the quarks have
the same flavor in the baryonium, Fig. 2 is of equal $N_c$
importance as Fig. 1.  The realistic situation is an interplay of
the two kinds of interactions described in Figs. 1 and 2.  The 
amplitude of the baryon-antibaryon scattering described in Fig. 2 
by assuming $N_f=1$ \cite{1} is proportional to $N_c$.  This 
divergent large $N_c$ behavior of the amplitude matches with 
that of baryon kinetic energies, and shows the strong interacting 
behavior between the baryon and the antibaryon.  

\begin{figure}
\includegraphics{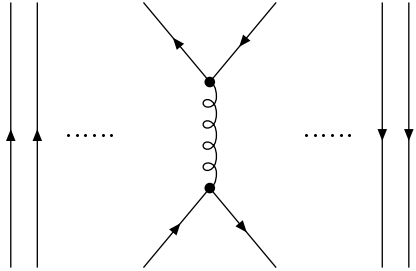}
\caption{\label{fig2}
Interaction inside a baryonium with a quark pair annihilation
\cite{1}}
\end{figure}

Assuming the existence of baryonia, interesting observation about 
hybrids and glueballs can be seen in the large $N_c$ limit.  In the 
case that all the quarks have the same flavor in a baryonium 
($N_f=1$), a hybrid state with valence content of $(N_c-1)$ quarks, 
$(N_c-1)$ antiquarks and one gluon, for instance, is large-$N_c$ 
enhanced due to the same reason of that baryon interaction is 
strong, as can be seen by cutting Fig. 2 in the middle.  This 
state and the baryonium transfer into each other constantly 
by the strong dynamics.  Therefore in fact, this hybrid state 
is physically not different from the baryonium in the large-$N_c$
limit.  For the same reason, this state can be equally identified
as being composed of $(N_c-2)$ quarks, $(N_c-2)$ antiquarks 
and two gluons.  Furthermore, and remarkably, a glueball 
composed of $N_c$ valence gluons is also in fact indistinguishable 
from the baryonium.  Referring to Fig. 2, it is easily seen that 
a valence gluon has a mass of $2$ valence quarks, which is 
about $2\Lambda_{\rm QCD}$.  The glueball composed of 
two valence gluons therefore has a mass about 
$4\Lambda_{\rm QCD}$.  It is clear that the glueball with 
$N_c$ valence gluons has a mass of the baryonium.  
We have stated that the baryonium, the hybrid and the glueball 
are the same state in $N_c=\infty$, $N_f=1$ QCD.  Although at 
the quark-gluon level this is not true, there is no way to 
distinguish them at the hadron level due to the confinement.  

On the other hand, above reasoning also implies that the existence 
of glueballs supports the existence of baryonia.  In the $N_f=0$ 
case, the confinement requires the existence of glueballs as hadrons.  
Light glueballs are massive with masses several times of 
$\Lambda_{\rm QCD}$.  Adding a single flavor into this case, lowest 
new hadrons include $\eta'$ meson due to chiral symmetry 
spontaneous breaking and anomaly, and $\Delta^{++}$ baryon.  In 
the case of $N_c\to\infty$, glueballs and baryonia 
$(\Delta^{++}, \bar{\Delta}^{--})$ with same quantum numbers are 
indistinguishable.  Baryonia are then generally expected.  The 
constituent quark mass is determined to be half of the constituent 
gluon mass.  With more flavor added, many new baryons with various 
flavor quantum numbers appear.  The baryonium existence beyond 
$N_f=1$ is less sound than the case of $N_f=1$.  

Note a glueball of $N_c$ valence gluons can transfer to a glueball of 
$(N_c-1)$ valence gluons.  But this transition rate is ${\mathcal O}(1)$.  
This is seen from Fig. 3.  The three-gluon vertex has a factor of 
$1/\sqrt{N_c}$.  There are $N_c$ ways to make the transition.  By 
considering the color quantum numbers of the gluons in Fig. 3, we know 
that the $N_c$ ways do not add coherently.  The rate of Fig. 3 is 
$1/N_c$.  Then the total transition rate from a $N_c$ valence gluon 
state into a $(N_c-1)$ valence gluon state is ${\mathcal O}(1)$.  It is 
$1/N_c$ suppressed compared to the transition rate of a baryonium into 
a one-gluon hybrid state.  Therefore the $N_c$ valence gluon state 
distinguishes itself from the $(N_c-1)$ valence gluon state.  The two 
states have ${\mathcal O}(1)$ mixing in amplitudes.  In other words, 
the number of valence gluons inside a glueball is well-defined in the 
large $N_c$ limit.  

\begin{figure}
\includegraphics{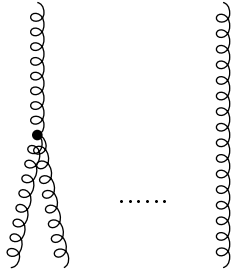}
\caption{\label{fig3}
$N_c$ valence gluons transfer to $N_c-1$ valence gluons}
\end{figure}

\section{Diquark-antidiquarks}

Unlike Ref. \cite{1}, we shall distinguish baryonium states and
diquark-antidiquark states.  Taking a color singlet quark pair
away from a baryonium, a color singlet 
$(N_c-1)$-quark-$(N_c-1)$-antiquark state can be always 
formed.  Their existence has been argued for in Ref. \cite{1}.  
The $(N_c-1)$ quarks form a ${\bf \bar{N}_c}$ representation, 
and $(N_c-1)$ antiquarks a ${\bf N_c}$ representation.  Such 
a state has a mass of about $2(N_c-1)\Lambda_{\rm QCD}$.  
It is large $N_c$ extension of the tetra-quark state, by taking 
$N_c=3$, this state is the diquark-antidiquark one.  In such a 
$(N_c-1)$-quark-$(N_c-1)$-antiquark system, if all the 
quark flavors are the same, the hybrid state of 
$(N_c-2)$-quark, $(N_c-2)$-antiquark and one gluon is 
large-$N_c$ enhanced and is physically not different from 
the $(N_c-1)$-quark-$(N_c-1)$-antiquark system.  

Let us consider in more detail the 
$(N_c-1)$-quark-$(N_c-1)$-antiquark state.  Taking $N_f=1$, 
processes similar to that described in Fig. 2 happen.  In most of 
the cases when the gluon is formed from quarks with different 
colors, interaction keeps the $(N_c-1)$ quarks in ${\bf \bar{N}_c}$ 
representation and $(N_c-1)$ antiquarks in ${\bf N_c}$ 
representation.  This process amplitude is proportional to $N_c$.  
When the gluon is formed from quarks with the same color, the 
final $(N_c-1)$ quarks generally do not stay in ${\bf \bar{N}_c}$ 
representation.  But this transition amplitude is ${\mathcal O}(1)$.  
The similar result can be obtained if we consider t-channel gluon 
exchanges.  Therefore, quark configuration of a state with the 
$(N_c-1)$ quarks in ${\bf \bar{N}_c}$ representation and 
$(N_c-1)$ antiquarks in ${\bf N_c}$ representation makes sense 
in the large $N_c$ limit.  In real situation ($N_c=3$), this is to 
say that ${\bf 3}\otimes\bf{\bar{3}}$ tetraquark configuration does 
not mix with ${\bf 6}\otimes\bf{\bar{6}}$ tetraquark configuration 
when $N_c=3$ is considered to be large.  

Further taking a color singlet quark pair away, a
$(N_c-2)$-quark-$(N_c-2)$-antiquark state is then formed 
with a mass being about $2(N_c-2)\Lambda_{\rm QCD}$.  
The $(N_c-2)$ quarks form a ${\bf \bar{N}_c\otimes\bar{N}_c}$ 
representation, and $(N_c-2)$ antiquarks a ${\bf N_c\otimes N_c}$ 
representation.  The situation is more complicated.

The above procedure might continue.  Finally, a valence
quark-antiquark state is formed with a mass being about 
$2\Lambda_{\rm QCD}$, which is just a meson.  ( Note that chiral 
symmetry breaking cannot be counted in this framework.)

In the same manner as we have discussed for the baryonium and 
the glueball in the last section, the single flavor $(N_c-1)$-quark
and $(N_c-1)$-antiquark system is not distinguishable from the
glueball composed of $N_c-1$ valence gluons.  And the two-quark
and two-antiquark system is indistinguishable from the glueball
composed of two valence gluons.

Once more flavors are included, above large $N_c$ consideration
becomes more complicated.  As an example, the $(N_c-1)$ quarks 
stay in the lowest energy state if they take $(N_c-1)$ different 
flavors.  In this case, the $(N_c-1)$ quark - $(N_c-1)$ antiquark state
transition to the hybrid state of $(N_c-2)$ quark, $(N_c-2)$ antiquark
and one gluon is $1/N_c$ suppressed compared to the single flavor
case.  They cannot be the same state in the large-$N_c$ limit, but
they have ${\mathcal O}(1)$ mixing in amplitudes.

\section{Decays and binding energies}

Decays of baryoniums and $(N_c-1)$ quark-$(N_c-1)$ antiquark states 
have been discussed in Ref. \cite{1}.  A baryonium decays into one meson 
and a $(N_c-1)$ quark-$(N_c-1)$ antiquark state.  A 
$(N_c-1)$ quark-$(N_c-1)$ antiquark state decays into one meson and a 
$(N_c-2)$ quark-$(N_c-2)$ antiquark state.  Such cascade decays continue 
until the final 4-quark state decays into two mesons.  These decays are slow.  

These decay rates are ${\mathcal O}(1)$.  This is easy to see from the fact 
that a color-singlet quark pair drops out of a baryonium or of a 
$(N_c-1)$ quark-$(N_c-1)$ antiquark state with an amplitude of 
${\mathcal O}(1)$.  

Theoretically, baryon-antibaryon systems might be 
difficult to deal with, because the typical energy of such 
interaction is proportional to $N_c$ \cite{1} which is the same 
$N_c$-dependence of baryon masses $\sim N_c\Lambda_{\rm QCD}$.  
However, as we have argued in Sect. II, it is expected that 
the baryon-antibaryon interacting energy is smaller than 
$N_c\Lambda_{\rm QCD}$ due to the confinement.  Furthermore, 
it is expected from large-$N_c$ QCD that the baryon-baryon 
typical interacting energy is also proportional to $N_c$ and 
phenomenologically, nuclear physics shows that the typical 
binding energy of a baryon inside a nuclear is only about a few  
MeV.  Therefore, we expect that the binding energy of a 
baryonium is actually a lot smaller than a baryon mass.  It makes 
molecular description (in the QCD sense) of baryonium states 
meaningful.  Such a system in fact can be well described by 
non-relativistic quantum mechanics.  Consequently, baryon spins 
decouple from the dynamics of baryoniums.  

To express the problem more clearly, in the large $N_c$-limit, 
baryon-baryon binding energy is $N_c\lambda$, and baryon-antibaryon 
binding energy is $N_c\lambda'$.  The confinement argument gives that 
$\lambda<\Lambda_{\rm QCD}$ and $\lambda'<\Lambda_{\rm QCD}$.  

The slow transition of a baryonium to a 
$(N_c-1)$ quark-$(N_c-1)$ antiquark state and one meson implies that 
the interacting strength inside the baryonium and the 
$(N_c-1)$ quark-$(N_c-1)$ antiquark state are the same.  In other 
words, in the large $N_c$ limit the constituent quark mass 
($\Lambda_{\rm QCD}$) can be taken the same in these two kinds of 
hadrons.

\section{Analysis}

We will make a numerical illustration through analyzing realistic 
situation.  In the semi-quantitative analysis, we only consider ground 
state hadrons.  Being accurate requires specification of meaning of 
$\Lambda_{\rm QCD}$ we have used.  This quantity describes 
the energy of an individual quark inside baryons.  In the following, 
we define $\bar{\Lambda}_{\rm QCD}$ to replace 
$\Lambda_{\rm QCD}$, 
\begin{equation}
M_{\rm ground~state~baryon}\equiv N_c\bar{\Lambda}_{\rm QCD}\,, 
\end{equation} 
namely $\bar{\Lambda}_{\rm QCD}$ is identified as a constituent 
quark mass in baryons.  Taking $N_c=3$, we have 
$\bar{\Lambda}_{\rm QCD}=(362\pm 50)$ MeV by taking the average 
of masses of a nucleon and $\Delta^{++}$.  

The analysis depends crucially on how much the baryonium binding 
energy is.  Different binding energy corresponds to different physical 
picture of hadrons.  As we will see it determines which group of 
hadron particles in the Particle Date Book is identified as 4-quark 
states.  We will mainly take a $10$ MeV binding energy.  The error 
of this $10$ MeV binding energy is hard to estimate.  Some other 
phenomenological works \cite{others} use about $300$ MeV binding 
energy.  A significantly larger binding energy will be considered 
briefly as a comparison later.  

The numerical analysis would be more appropriate to the $N_f=1$ 
case, however, it will go beyond that to $N_f=2$ and $3$ without 
further mention of that the latter cases are more assumption-dependent.  

\subsection{$10$ MeV binding energy}  

We take the binding energy to be $10$ MeV, that means that 
$\lambda'\simeq\lambda$ or a little bit larger, considering 
both $\lambda$ and $\lambda'$ are essentially determined by 
$\Lambda_{\rm QCD}$ and the confinement.  Nuclear physics 
tells us that $N_c\lambda$ is about a few MeV.  We expect 
that $N_c\lambda'\simeq 10$ MeV typically .  A recent study 
from the Skyrmeon model shows that the baryonium binding 
energy is indeed about $10$ MeV \cite{yan}.

Consider the case of only one flavor, the lowest baryonium is 
s-wave ($\Delta^{++}$ $\bar{\Delta}^{--}$), with the quark 
being the up-quark as an example, its mass is about 
$2M_{\Delta^{++}}-10 {\rm MeV} \simeq 2450$ MeV.  As 
we have argued, this state can be also identified as a $0^{-+}$ 
ground state of three-gluon glueball in the large $N_c$ limit.  
The infered constituent gluon mass is consistent with those 
via other methods \cite{other}.  The actual glueball mass maybe 
a bit lower than the above value, because the actual flavor 
number is more than one.  With one more light flavor being 
introduced, numerically it is estimated that the mass of the 
lowest proton-antiproton molecular state is about $2M_N-10$ 
MeV $\simeq 1866$ MeV.  This molecular state mixes with 
the $0^{-+}$ glueball.  In the large $N_c$ limit, this mixing 
depends on large $N_c$ generalization of the nucleon, which 
we do not consider in this paper.  

Experimentally, BES collaboration has found two baryonium
candidates, $X(1860)$ \cite{7} and $X(1835)$ \cite{8}.  
They were then theoretically studied \cite{b,yan}.  Considering
the uncertainties of the binding energy, these states are
consistent with our expectation.  Furthermore, the corresponding 
$0^{-+}$ state exists due to the different baryon 
spin combination.  Their approximate degeneracy is a result 
of baryon spin decoupling.  As a check, $p\bar{n}$,
$n\bar{p}$ and $n\bar{n}$ states should have a degenerate 
mass as $p\bar{p}$ which is about $1835$ MeV or $1860$ 
MeV.  In the three light flavor case, the lowest baryonium 
$(p,\bar{\Lambda})$ or $(n,\bar{\Lambda})$ is expected to 
have a mass of $M_N+M_\Lambda-10$ MeV $\simeq 2045$ 
MeV and the baryonium $(\Lambda,\bar{\Lambda})$ with a 
mass of $2M_{\Lambda}-10 {\rm MeV} \simeq 2220$ MeV.  
$(p,\bar{\Lambda})$ is consistent with the experiment 
$\sim 2075\pm 13$ MeV \cite{bes}.  

A $(N_c-1)$ quark-$(N_c-1)$ antiquark state is of a mass about 
$2(N_c-1)\bar{\Lambda}_{\rm QCD}$.  From the argument of last 
section, it is reasonable to take the constituent quark mass in a 
tetra-quark state to be the same as that in a baryonium.  In the case 
of only one flavor, the 4-quark state is $(uu~\bar{u}\bar{u})$.  Its 
lowest mass is estimated to be $(2M_{\Delta^{++}}-10 {\rm MeV})
-2\bar{\Lambda}_{\rm QCD} \simeq 1740\pm 100$ MeV.  
This state can be also regarded as a $(u\bar{u}g)$ hybrid or 
a $0^{++}$ ground state two-gluon glueball.  It should be 
identified as $f_0(1710)$ in our scheme.   Our estimation is 
consistent with lattice calculation \cite{lattice}. 

With one more flavor included, the lowest mass can be written as
\begin{equation}
2(N_c-1)\bar{\Lambda}_{\rm QCD}\simeq 
[M_{X(1835)}~{\rm or}~M_{X(1860)}] -2\bar{\Lambda}_{\rm QCD}
\simeq (1110-1140)\pm 100 {\rm MeV}\,.
\end{equation}
This is the 4-quark ground state $(ud~\bar{u}\bar{d})$ with 
both spin and isospin $0$.  It can be identified as $f_0(1370)$ 
which has a mass ranging from $1200$ to $1500$ MeV \cite{9}.  
As being noted, it mixes with the hybrid state $(q\bar{q}g)$ 
with $q$ standing for the $u$- or $d$-quark.  

As we have seen that 4-quark state mass estimation has an 
uncertainty $\sim 200$ MeV, above numbers are of limited use 
practically.  Our point is that within the uncertainty, there should be 
4-quark states, and they indeed have experimental correspondence.  
After the strange quark is introduced, three kinds of lowest
diquarks can be formed, $ud$, $us$ and $ds$.  The 4-quark states
form a nonet.  They were studied by many authors 
\cite{di,cheng,14,m,maiani}.  
In our work, their mass differences are expected to be determined 
by the strange quark mass, which do not subject to the large 
uncertainty of large $N_c$ approximation.  
So the ground 4-quark states are naturally identified as 
$f_0(1370)$, $a_0(1450)$, $K_0^*(1430)$, $f_0(1500)$; 
more explicitly, 
$f_0(1370)(ud\bar{u}\bar{d})$, $K_0^*(1430)(ud\bar{u}\bar{s},
ud\bar{d}\bar{s}, us\bar{u}\bar{d}, ds\bar{u}\bar{d})$,
$a_0(1450)(us\bar{d}\bar{s}, ds\bar{u}\bar{s})$,
$a_0(1450)(s(n\bar{n})_{-}\bar{s})$, 
$f_0(1500)(s(n\bar{n})_{+}\bar{s})$,
where $(n\bar{n})_{\pm}\equiv (u\bar{u}\pm d\bar{d})/\sqrt{2}$.  
The mixing among the $f_0$ states are an ${\mathcal O}(1)$ effect.  
Note that our 4-quark state identification is different from 
most of previous studies \cite{di,cheng,14,m} where it is 
$\sigma$, $\kappa$, $a_0(980)$ and $f_0(980)$ that 
are taken to be the lowest $0^{++}$ 4-quark states.  

\subsection{Large binding energy}  

As a comparison, let us consider the case of large baryon-antibaryon 
interacting energies.  We know that $\lambda\simeq\lambda'$ is still 
an assumption.  If the annihilation effect is important, $\lambda'$ is 
possibly a lot larger.  Refs. \cite{others} took it to be about $200$ 
MeV which still makes the molecular picture of baryoniums sensible.  
Now we fix $N_c\lambda'$ by requiring that 4-quark ground states 
correspond to $\sigma$, $\kappa$, $a_0(980)$ and $f_0(980)$.  For 
a large binding energy,  the mass of a diquark-antidiquarks is written 
as $2(N_c-1)\bar{\Lambda}_{\rm QCD}''$, while the mass of a 
baryonium is $2N_c\bar{\Lambda}_{\rm QCD}'$.  It is reasonable for 
ground states that 
$\bar{\Lambda}_{\rm QCD}''\simeq\bar{\Lambda}_{\rm QCD}'$  
in the large-$N_c$ limit, because we can imagine a dynamical 
${\mathcal O}(1)$ process to generate a ground state diquark-antidiquark 
from a ground state baryonium via emitting a meson with the mass 
being $2\bar{\Lambda}_{\rm QCD}$.  The interaction energy 
between a diquark-antidiquark and a meson is $1/N_c$ 
suppressed compared to $N_c\lambda'$, and will be neglected, 
as we have also implicitly done in last subsection.  
Taking the strange quark mass $m_s=150$ MeV, from 
$M_{f_0(980)}\simeq 2(N_c-1)\bar{\Lambda}_{\rm QCD}'+2m_s$, 
we obtain that $\bar{\Lambda}_{\rm QCD}'\simeq 170$ MeV.  
In this case, $M_{\sigma}\simeq 680$ MeV.  Lowest $0^{-+}$ 
baryoniums should be then about $1020$ MeV which means a 
$960$ MeV binding energy.  However, there is no such 
baryonium correspondence in the Particle Data Book actually.  

Therefore, considering practical situation, large-$N_c$ QCD 
analyses prefer a small baryonium binding energy.  In the 
following we will not consider the large binding energy case.  

\subsection{Decay widths}

A baryonium decays into one meson and one tetra-quark state, and a 
tetra-quark state into two mesons \cite{1}.  The decay rates are 
${\mathcal O}(1)$.  Therefore, from the large $N_c$ point of view, 
diquark-antidiquark states decay slowly, and this also makes them 
distinguishable from two meson states.  The $0^{-+}$ baryonium 
$X(1835)$ or $X(1860)$ decays in p-wave into $f_0(1370)$ and 
$\sigma$.  Note that it cannot decay to s-wave $f_0(1370)$ and 
$\eta'$ due to the phase space.  Therefore the dominant decay products 
are $\rho\rho\pi\pi$.  This can be checked by future experiments.  

\section{Heavy hadrons}

Now we consider the case of inclusion of a single heavy quark.
Heavy quark effective theory (HQET) \cite{12} provides a
systematic way to investigate hadrons containing a single heavy
quark.  It is an effective field theory of QCD for such heavy
hadrons.  In the limit $m_Q/\Lambda_{\rm QCD}\to\infty$, the heavy
quark spin-flavor symmetry is explicit.  The hadron mass is
expanded as
\begin{equation}
M_H = m_Q+\bar{\Lambda}_H+{\mathcal O}(1/m_Q)\;.
\end{equation}
To obtain the HQET defined, universal heavy hadron mass
$\bar{\Lambda}_H$, however, some non-perturbative QCD methods 
have to be used.  We can apply the large $N_c$ method.  Heavy 
baryons were studied via this method \cite{w,c,l,10a,l1}.  Let us 
first consider the relation of the quantity $\bar{\Lambda}_H$ of a 
ground state heavy baryon and the nucleon mass.  Heavy baryons 
contain $(N_c -1)$ light quarks, and one "massless" heavy quark 
(modular $m_Q$).  The mass or the energy of the baryon is determined 
by the summation of the energies of individual quarks.  The kinetic 
energy of the heavy quark is typically $\bar{\Lambda}_{\rm QCD}$ 
like that of the light quark.  The interaction energy between 
the heavy quark and any of the light quarks is typically 
$\bar{\Lambda}_{\rm QCD}/{\rm N_c}$.  So the interaction energy 
between the heavy quark and the whole light quark system scales as 
$\bar{\Lambda}_{\rm QCD}$.  However, the total interaction energy of 
the light quark system itself scales as $N_c\bar{\Lambda}_{\rm QCD}$.  
Therefore in the large $N_c$ limit, $\bar{\Lambda}_H=M_N$ 
where the uncertainty of the equation is  
${\mathcal O}(1)\sim\bar{\Lambda}_{\rm QCD}$ in $1/N_c$ expansion.

Actually in the mass relation between heavy baryons and corresponding 
light baryons under the large $N_c$ limit, the uncertainty is smaller 
than $\bar{\Lambda}_{\rm QCD}$.  This is because the heavy quark 
constituent mass (modular $m_Q$) does not deviate from 
$\bar{\Lambda}_{\rm QCD}$ very much.  For an example, the 
$\Lambda_Q$ baryon mass $\bar{\Lambda}_{\Lambda_Q}$ is about 
$0.80$ GeV \cite{13}.  It is more reasonable to take 
$M_N-\bar{\Lambda}_{\Lambda_Q}\sim 0.15$ GeV to be the 
uncertainty in the following analysis.  

Heavy baryoniums containing a heavy quark are analyzed in the same 
large-$N_c$ spirit of the light quark case.  The $0^-$ ground state 
baryoniums $(\Lambda_c, \bar{N})$ and $(\Lambda_c, \bar{\Lambda})$ 
have masses
\begin{equation}
\begin{array}{lll}
M_{(\Lambda_c, \bar{N})}&\simeq&
m_c-m_s+M_{(N,\bar{\Lambda})}\simeq 3.33\pm 0.15~{\rm GeV}\,,\\
M_{(\Lambda_c, \bar{\Lambda})}&\simeq&
m_c-m_s+M_{(\Lambda,\bar{\Lambda})}\simeq 3.50\pm 0.15
~{\rm GeV}\,,
\end{array}
\end{equation}
where $m_c$ is taken to be $1.43$ GeV \cite{13}.  This is consistent 
with naive estimation $M_{(\Lambda_c, \bar{N})}=
M_{\Lambda_c}+M_N-10~{\rm MeV}\simeq 3.21$ GeV and 
$M_{(\Lambda_c, \bar{\Lambda})}=
M_{\Lambda_c}+M_{\Lambda}-10~{\rm MeV}\simeq 3.40$ GeV.  
They also have corresponding degenerate $1^-$ states due to baryon spin 
decoupling.    

For heavy diquarks, because of the heavy quark spin symmetry, existence 
of a spin-zero diquark $Qq$ implies that a $Qq$ spin-one diquark also 
exists.  This point was also noticed in Ref. \cite{m}.  The spectrum of the 
lowest 4-quark charm states is
\begin{equation}
\begin{array}{lll}
M_{(cd, \bar{u}\bar{d})}&=&M_{(cu, \bar{u}\bar{d})}=m_c+M_{f_0(1370)}
\simeq (2.54-2.57)\pm 0.15~{\rm GeV} \,, \\
M_{(cd, \bar{u}\bar{s})}&=&M_{(cu, \bar{u}\bar{s})}=
M_{(cd, \bar{d}\bar{s})}=M_{(cu, \bar{d}\bar{s})}=
M_{(cs, \bar{u}\bar{d})} \\
&=&m_c+M_{a_0(1450)}\simeq 2.84\pm 0.15~{\rm GeV} \,, \\
M_{(cs, \bar{u}\bar{s})}&=&M_{(cs, \bar{d}\bar{s})}
=m_c+M_{f_0(1500)}\simeq 2.94\pm 0.15~{\rm GeV} \,. \\
\end{array}
\end{equation}
In the heavy quark limit, we have the degeneracy 
of $0^+$, $1^+$ and $2^+$ 4-quark states.  

Therefore, we expect a rich charm hadron spectrum ranging from 
$2.54$ GeV to $3.50$ GeV.  The $1/m_Q$ uncertainty is about
$\Lambda_{\rm QCD}^2/m_c\sim 60$ MeV.  The bottom case is 
the same except for smaller $1/m_Q$ correction 
because of the heavy quark flavor symmetry,
\begin{equation}
\begin{array}{lll}
M_{(\Lambda_b, \bar{N})}&\simeq&
m_b-m_s+M_{(\Lambda,\bar{N})}\simeq 6.73\pm 0.15~{\rm GeV}\,,\\
M_{(\Lambda_b, \bar{\Lambda})}&\simeq&
m_b-m_s+M_{(\Lambda,\bar{\Lambda})}\simeq 6.90\pm 0.15
~{\rm GeV}\,, \\
M_{(bd, \bar{u}\bar{d})}&=&M_{(bu, \bar{u}\bar{d})}
\simeq m_b+M_{f_0(1370)}\simeq (5.94-5.97)\pm 0.15~{\rm GeV} \,, \\
M_{(bd, \bar{u}\bar{s})}&=&M_{(bu, \bar{u}\bar{s})}=
M_{(bd, \bar{d}\bar{s})}=M_{(bu, \bar{d}\bar{s})}=
M_{(bs, \bar{u}\bar{d})} \\
&\simeq&m_b+M_{a_0(1450)}\simeq 6.24\pm 0.15~{\rm GeV} \,, \\
M_{(bs, \bar{u}\bar{s})}&=&M_{(bs, \bar{d}\bar{s})}
\simeq m_b+M_{f_0(1500)}\simeq 6.34\pm 0.15~{\rm GeV} \,, \\
\end{array}
\end{equation}
where $m_b\simeq 4.83$ GeV \cite{13}.  $M_{(\Lambda_b, \bar{N})}$ 
and $M_{(\Lambda_b, \bar{\Lambda})}$ are consistent with 
$M_{\Lambda_b}+M_n-10~{\rm MeV}\simeq 6.61$ GeV and 
$M_{\Lambda_b}+M_{\Lambda}-10~{\rm MeV}\simeq 6.80$ GeV, 
respectively.  

For hadrons containing a pair of heavy quarks, the two heavy
quarks intend to combine into a tighter object which is described
by non-relativistic QCD.  However, if the two heavy quarks are
separated by $1/\Lambda_{\rm QCD}$ or more in certain hadrons, 
the above HQET procedure can be applied to these hadrons.  For 
examples, the following lowest baryoniums and 4-quark states may exist,
\begin{equation}
\begin{array}{lll}
M_{(\Lambda_c, \bar{\Lambda}_c)}&\simeq&
2m_c-2m_s+M_{(\Lambda,\bar{\Lambda})}
\simeq 4.78\pm 0.15~{\rm GeV}\,, \\
M_{(cu, \bar{c}\bar{u})}&=&M_{(cu, \bar{c}\bar{d})}=
M_{(cd, \bar{c}\bar{u})}=M_{(cd, \bar{c}\bar{d})}\simeq 2m_c
+M_{f_0(1370)}\simeq (3.97-4.00)\pm 0.15~{\rm GeV} \,, \\
M_{(cu, \bar{c}\bar{s})}&=&M_{(cd, \bar{c}\bar{s})}=
M_{(cs, \bar{c}\bar{u})}=M_{(cs, \bar{c}\bar{d})}
\simeq 2m_c+M_{a_0(1450)}\simeq 4.27\pm 0.15~{\rm GeV} \,, \\
M_{(cs, \bar{c}\bar{s})}&\simeq&2m_c+M_{f_0(1500)}\simeq 4.37
\pm 0.15~{\rm GeV}\,,
\end{array}
\end{equation}
where $M_{(\Lambda_c, \bar{\Lambda}_c)}$ is consistent with 
$2M_{\Lambda_c}-10~{\rm MeV}\simeq 4.83$ GeV, and the 
uncertainty due to $1/m_Q$ effects is 
$\Lambda_{\rm QCD}^2/2m_c\simeq 30$ MeV.
The state $(cs, \bar{c}\bar{s})$ is consistent with $Y(4260)$
\cite{14} which cannot be identified as $(\Lambda_c,
\bar{\Lambda}_c)$ \cite{15} in our scheme.  
Considering $1/N_c$ uncertainties, $X(3940)$ can be 
($cq~\bar{c}\bar{q}$) ($q=u,d$).  In that case, charged 
($cu~\bar{c}\bar{d}$) state is also expected around $3940$ MeV.  

\section{Summary and discussion}

From the large $N_c$ QCD point of view, we have considered
baryoniums, four-quark states, hybrids and glueballs.  The existence of 
baryonium states is argued for from existence of nuclei. These hadrons 
are constructed from baryons.  We have argued that in $N_f=1$ large 
$N_c$ QCD, a baryonium is always identical to a glueball with 
$N_c$ valence gluons.  $f_0(1370)$, $a_0(1450)$, 
$K^*_0(1430)$ and $f_0(1500)$ are identified as the lowest four-quark
nonet.  The glueball with three valence gluons has a mass about $2450$
MeV.  $f_0(1710)$ is identified as the glueball with two valence gluons.
Combining with HQET, we have predicted heavy baryoniums and heavy 
four-quark states.

This work can be viewed as a large $N_c$ QCD extension of the 
Fermi-Yang model.  A classification of hadrons is given a large $N_c$ 
QCD basis.  We have constructed hadron spectra from baryoniums,
because our starting point is baryons.  The reversed procedure is
not necessarily true.  For an example, the one-gluon hybrid state
existing in this scheme can always be generated from or identified
as a diquark-antidiquark state.  
The large $N_c$ QCD arguments and consequent estimation help us 
understanding relevant experimental results.  However, they do not 
result in a precise mathematical description for the hadrons we have 
studied in this paper.  Finding such a systematic description is one of 
the task in sloving nonperturbative QCD.  

The uncertainties of the analysis should be discussed.  Of course, 
existence of baryoniums as well as the $10$ MeV binding energy is still 
an assumption, but it is supported by the large $N_c$ analysis.  Even if 
$N_c=3$, baryoniums are still expected.  In many cases, qualitative 
conclusion of large $N_c$ QCD is also true when $N_c=3$.  The 
baryon-baryon strong interaction in large $N_c$ QCD implies existence 
of baryon bound states.  Indeed in real QCD with $N_c=3$, nuclei  
exist.  The meson-meson interaction is vanishing in the large 
$N_c$-limit, therefore, there is no molecular states of mesons.  And 
in the real world, meson molecular states seem non-existent.  

It is important to discuss the $1/N_c$ corrections to our numerical 
analysis.  The estimated masses of the baryoniums and 
diquark-antidiquark states would have ${\mathcal O}(1)\sim (200-300)$ 
MeV uncertainties.  But the relative masses of the above hadrons have 
no that large uncertainties.  For an example, once the binding energy of 
$p\bar{p}$ is fixed as $10$ MeV, then the $\Lambda\bar{\Lambda}$ 
binding energy is $10$ MeV with an uncertainty of about $30\%$ due 
to SU(3) violation.  Namely, the baryoniums $p\bar{p}$ and 
$\Lambda\bar{\Lambda}$ mass difference does not subject to large 
$1/N_c$ corrections.  More accurate treatment of baryoniums can be 
similar to that in Refs. \cite{10,10a} by taking account baryonium binding 
energies.  Furthermore, for the diquark-antidiquark states, their mass 
differences similarily have no large $1/N_c$ uncertainties.  
For another example, mass differences of states 
$(cd, \bar{u}\bar{d})$, $(cd, \bar{u}\bar{s})$ and $(cs, \bar{u}\bar{s})$ 
do not subject to $\bar{\Lambda}_{\rm QCD}$ uncertainty.  
The discovery of the baryoniums and the diquark-antidiquark states with a 
single heavy quark and their mass estimation given in Eqs. (3-5) will be 
tests of our understanding in the near future.

\begin{acknowledgments}
I would like to thank Yuan-Ben Dai, Hai-Yang Cheng, Ming-Xing Luo, 
Shi-Lin Zhu, Yu-Qi Chen, Jian-Ping Ma and Yue-Liang Wu for helpful 
discussions.  This work was partially finished during my visits to Academia 
Sinica in Taipei and Zhejiang University in Hangzhou.  This work was 
supported in part by the National Science Foundation of China
(Nos. 10675154) and CAS.
\end{acknowledgments}

\end{document}